\documentclass[5p,twocolumn,times,number]{elsarticle}

\usepackage{graphicx}
\usepackage{amsmath}   
\usepackage{hyperref}
\usepackage{amsthm}
\usepackage{lineno}

\begin{document}

\begin{frontmatter}

\title{Na-based crystal scintillators for next-generation rare event searches}

\author[a]{S.~Nagorny}
\author[b,c]{C.~Rusconi\corref{cor}}
\ead{Claudia.Rusconi@LNGS.INFN.IT}
\author[d,e]{S.~Sorbino}
\author[f]{J.W.~Beeman}
\author[d,e]{F.~Bellini}
\author[e]{L.~Cardani}
\author[g]{V.D.~Grigorieva}
\author[h,i]{L.~Pagnanini}
\author[c]{S.~Nisi}
\author[c]{S.~Pirro}
\author[l]{K.~Sch\"affner}
\author[g,m]{V.N.~Shlegel}
\cortext[cor]{Corresponding author}

\address[a]{Queen's University, Physics Department, K7L 3N6, Kingston (ON), Canada}
\address[b]{Department of Physics and Astronomy, University of South Carolina, Columbia, SC 29208 - USA}
\address[c]{INFN - Laboratori Nazionali del Gran Sasso, 67010 Assergi - Italy}
\address[d]{Dipartimento di Fisica, Sapienza Universit\`a di Roma, P.le Aldo Moro 2, 00185, Rome, Italy}
\address[e]{INFN - Sezione di Roma 1, 00185 Roma - Italy}
\address[f]{Lawrence Berkeley National Laboratory , Berkeley, California 94720, USA}
\address[g]{Nikolaev Institute of Inorganic Chemistry, 630090 Novosibirsk - Russia}
\address[h]{Dipartimento di Fisica, Universit\`{a} di Milano - Bicocca, 20126 Milano - Italy}
\address[i]{INFN - Sezione di Milano Bicocca, 20126  Milano - Italy}
\address[l]{Max-Planck-Institut f\"ur Physik, 80805 M\"unchen - Germany}

\address[m]{Novosibirsk State Tech University, 20 Karl Marx Prospect, 630092 Novosibirsk - Russia}

\begin{abstract}
The growing interest in clarifying the controversial situation in the Dark Matter sector has driven the experimental efforts towards new ways to investigate the long-standing DAMA/LIBRA result. Among them, low-temperature calorimeters based on Na-containing scintillating crystals offer the possibility to clarify the nature of the measured signal via particle identification. Here we report the first measurement of Na-containing crystals, based on material different from NaI, i.e. Na$_2$Mo$_2$O$_7$ and Na$_2$W$_2$O$_7$, pointing out their excellent performance in term of energy resolution, light yield, and particle identification.
\end{abstract}

\end{frontmatter}

\section{Introduction}
Several astronomical observations have shown that the visible and ordinary matter represents only a small fraction of the Universe, a large part of which is expected to be in the form of non-visible components, namely Dark Matter (DM), and Dark Energy \cite{Plank:2015}. 

While the Cosmology seems to corroborate this hypothesis, direct DM searches depict a controversial scenario. The DAMA/LIBRA collaboration, operating 250 kg of NaI(Tl) scintillators, measures a signal compatible with the presence of DM in our galactic halo \cite{Bernabei:2018xsa}. The other experiments in this sector, even if their sensitivity is such as to measure the DAMA/LIBRA signal, do not observe it.

This comparison among experiments based on different target materials, relies on the so-called {\it standard assumptions} about the DM halo distribution and the DM-nucleus cross-section, whose dependence on the atomic mass and the nuclear form factor is unknown. The first step to solve the puzzle is a model-independent cross-check of the DAMA/LIBRA signal, which is feasible only using the same target material.

I-based experiments (KIMS \cite{Kim:2012rza}, PICO \cite{Fushimi:2015sew}) have already ruled out the participation of iodine in the measured DAMA/LIBRA signal, while NaI-based experiments are currently in operation (ANAIS \cite{Amare:2019PRL}, COSINE \cite{Adhikari:2019PRL}) or construction (SABRE \cite{Antonello:2018fvx}). These experiments exploit the same technique as DAMA/LIBRA, as they measure the scintillation light produced by particle interactions in NaI(Tl) crystals. 

Recently, the COSINUS collaboration has demonstrated that NaI and CsI crystals can also be operated as low-temperature calorimeters \cite{Angloher:2017sft,Angloher:2016hbv}. Taking advantage from the lower energy threshold, and the particle identification capability \cite{Angloher:2016ooq}, low-temperature calorimeters have promising prospects to give new insights on the long-standing DAMA/LIBRA claim.

The same technique can also be used with other Na-containing crystals, in order to overcome some technical difficulties related to the hygroscopicity of NaI.

In this paper we investigate for the first time two Na-containing crystals, Na$_2$Mo$_2$O$_7$ and Na$_2$W$_2$O$_7$, and prove that they can be operated as low-temperature calorimeters obtaining excellent performance term of energy resolution, light yield, and particle discrimination.

\section{Cryogenic calorimeters}
Cryogenic calorimeters, historically also called bolometers, are one of the most promising technologies for the search of rare events~\cite{Pirro:2017ecr} such as neutrinoless double beta decay (CUORE~\cite{Alduino:2017ehq}, CUPID-0~\cite{Azzolini:2019tta,Azzolini:2018oph,Azzolini:2018yye}, CUPID-Mo~\cite{Armengaud:2017hit,Armengaud:2019loe}, AMoRE~\cite{Alenkov:2019jis}) and dark matter interactions (CRESST~\cite{Abdelhameed:2019hmk}, COSINUS~\cite{Angloher:2017sft}, EDELWEISS~\cite{Armengaud:2018cuy}, SuperCDMS~\cite{Agnese:2018col}).

The working principle of these devices is rather simple: energy deposits into an absorber give rise to temperature variations, which can be converted into electrical signals through dedicated sensors.

Common absorbers are dielectric and diamagnetic crystals, for which we can assume a lattice-dominated specific heat:
\begin{equation}
    c(T) = \frac{12}{5}\pi^4N_Ak_B\left(\frac{T}{\Theta_D}\right)^3
\end{equation}
where N$_A$ is the Avogadro Number, k$_B$ the Boltzmann constant and $\Theta_D$ the Debye temperature. 

If the crystal is cooled at cryogenic temperatures ($\approx$10\,mK) the thermal capacitance $C$ can reach values as low as 10$^{-9}$-10$^{-10}$\,J/K (depending on the exact values of $\Theta_D$ and of the crystal mass). Such a small capacitance allows to measure temperature variation up to a few hundreds of $\mu$K for an energy deposit of 1\,MeV in the crystal.

The temperature variations are then converted into sizable voltage signals exploiting dedicated sensors. In this work, we used Neutron Transmutation Doped Germanium thermistors (NTD) attached to the crystals. Typical voltage signals produced by NTDs ranges from hundreds to thousands of $\mu$V/MeV.
The light detectors (LD) used with this technique are themselves  bolometers~\cite{Beeman:2013zva}: they usually  consists of a thin  semiconductor (dark)  crystal wafer  that absorbs the scintillation photons giving rise to a sizable heat pulse that is proportional to the number of absorbed photons.

The technique of cryogenic calorimeters offers an exquisite energy resolution (usually around 0.1$\%$) and versatility in the choice of the compounds, enabling the investigation of different materials. Nevertheless, the crystals operated as absorbers must feature good phonons propagation and excellent radio-purity.
Moreover, crystals emitting a large amount of light at cryogenic temperatures are preferable, as they allow the simultaneous measurement of the heat signal and of the emitted light, enabling particle identification.

\section{Crystal production}
The crystals characterized in this work were produced at Nikolaev Institute of Inorganic Chemistry, Siberian Branch, Russian Academy of Science, at Novosibirsk, Russia, using the Low Temperature Gradient Czochralski (LTG-Cz) technique operating three-zone furnace~\cite{Atuchin:2011,Vasiliev:2011,Spassky2012}. 

The distinctive feature of this method is a temperature gradient of 0.1–1.0 K/cm in the melt, much lower in comparison to the conventional Czochralski technique in which the temperature gradient is around 10–100 K/cm (see~\cite{BOROVLEV2001305} for more details). The second important feature of this technique is the use of closed platinum crucibles. Combination of these two features leads to the  minimization of charge components evaporation from the melt, and losses of the initial materials do not exceed 0.5$\%$.

Standard market 4N purity grade  Na$_2$CO$_3$ and WO$_3$ powders were used as starting materials for the growth of  Na$_2$W$_2$O$_7$ as  described in details in~\cite{gavrilova2014}.

\begin{figure}[tb]
\centering
\includegraphics[clip=true,width=0.48\textwidth]{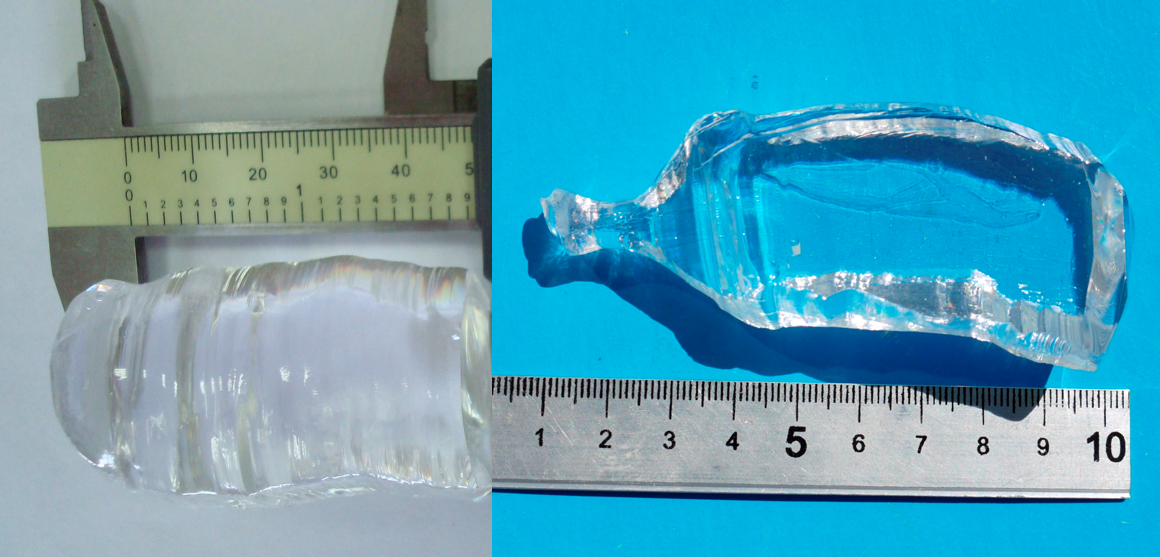}
\hfill
\caption{Picture of the Na$_2$W$_2$O$_7$ (left) and Na$_2$Mo$_2$O$_7$ (right) crystalline boules.}
\label{fig:boules} 
\end{figure}

The stoichiometric mix of  Na$_{2}$CO$_{3}$ and WO$_{3}$ powders was used as starting materials for the charge preparation: they were allowed to react at 650 $^{\circ}$C for 12 h into the platinum crucible according to the reaction: Na$_{2}$CO$_{3}$ + 2WO$_{3}$ $\rightarrow$ Na$_{2}$W$_{2}$O$_{7}$ + CO$_{2} \uparrow$. 
The obtained Na$_{2}$W$_{2}$O$_{7}$ powder  was then  further annealed  at 750 $^{\circ}$C for 12 hours in the same crucible and in air atmosphere  to yield the final charge.
Finally the crystal was pulled along [001] axis at 1.5–2.5 mm/h rate and rotation rate 3–12 rpm, respectively. A colorless, transparent and non-hygroscopic Na$_{2}$W$_{2}$O$_{7}$ single crystal with excellent phase purity was obtained~\cite{gavrilova2014} as shown in  Fig.~\ref{fig:boules}(left).

Deeply purified molybdenum oxide (MoO$_{3}$)~\cite{Shlegel:2013iga,Grigorieva:2018qaz} and commercial sodium carbonate Na$_{2}$CO$_{3}$ (4N purity grade) were used for the solid-state synthesis of Na$_{2}$Mo$_{2}$O$_{7}$ powder according to the reaction: 2MoO$_{3}$ + Na$_{2}$CO$_{3}$ $\rightarrow$ Na$_{2}$Mo$_{2}$O$_{7}$ + CO$_{2} \uparrow$.
The synthesis was carried out in a platinum crucible with dimensions 70 mm (dia) and 130 mm (length). 
The compound was heated up to 350 $^{\circ}$C with the rate of 100 $^{\circ}$C/h, then heated up to 650 $^{\circ}$C with the rate 20 $^{\circ}$C/h and held at this temperature for 5 h for homogenization of the melt.
The Na$_{2}$Mo$_{2}$O$_{7}$ crystal was then  grown in the same crucible and  pulled  along [010] axis with 0.3–2.5 mm/h rate and rotation rate 3–12 rpm, respectively. A high optical quality, colorless Na$_{2}$Mo$_{2}$O$_{7}$ single crystal was obtained \cite{Grigorieva:2019} and is presented in Fig.~\ref{fig:boules}(right). 

Both crystals, disodium molybdate and disodium tungstate, were grown in air atmosphere.

Finally, two 10$\times$10$\times$10 mm$^3$ samples of Na$_2$Mo$_2$O$_7$ and Na$_2$W$_2$O$_7$ crystals were cut from the single crystalline boules for characterization as scintillating bolometers. A sample of each crystal was then tested for internal contamination using a  High Resolution Inductively
Coupled Plasma-Mass Spectrometer (HR-ICP-MS, Thermo Fisher Scientific ELEMENT2) at the Laboratori Nazionali del Gran Sasso of INFN (LNGS, Italy). The results of the analyses are listed in Tab.~\ref{Table:contaminations}.

\begin{table}[!hbtp]
\centering
\begin{tabular}{lcc}
Element & Na$_{2}$Mo$_{2}$O$_{7}$ & Na$_{2}$W$_{2}$O$_{7}$\\

\hline
\hline
$	K	$&$	<15 	$&$	<10 	$\\
$	Cr	$&$	<.02	$&$	<.02	$\\
$	Mn	$&$	<.03	$&$	<.03	$\\
$	Fe	$&$	<5	    $&$	<5	$\\
$	Ni	$&$	<.05	$&$	<.05	$\\
$	Zn	$&$	<.02	$&$	<.02	$\\
$	Cd	$&$	<20 	$&$	\textbf{2.5	}$   \\
$	Sn	$&$	<2	    $&$	<2	$\\
$	Sb	$&$	<.02	$&$	<.02	$\\
$	Pb	$&$	\textbf{.03	}    $&$\textbf{	.02}	$\\
$	Th	$&$	<.02	$&$	<.1	$\\
$	U	$&$	<.02	$&$	<.03	$\\

\hline
\hline
\end{tabular}
\caption{Contaminations in ppm obtained with ICPMS on some common  
metals. Samples were dissolved in a 1:5 solution of HF and HNO$_3$ at  $75^ \circ$C with 
dilution factors of 470 (Na$_{2}$Mo$_{2}$O$_{7})$ and 680 (Na$_{2}$W$_{2}$O$_{7}$).}
\label{Table:contaminations}
\end{table}

\section{Experimental setup}
The mounting hosting the crystals consists of a copper basement covered with a sheet of a plastic reflector (Vikuiti$^{TM}$) on which the crystals rest without any holding structure. Moreover, the plastic reflector tightly surrounds the lateral faces of each crystal, to maximize the light collection while keeping them in position.

Each crystal is equipped with a $2.85\times2\times0.5$ mm$^3$ NTD through six dots of Araldit Rapid$^{TM}$ glue ($\sim$0.05 mm height, $\sim$0.7 mm diameter). The NTD thermistors are provided with 50 $\mu$m thick gold wires for the sensor readout, which are connected to electrically insulated copper pins held in a copper frame surrounding the mounting. The pins are then connected to the cryostat readout through copper wires.

\begin{figure}[tb]
\centering
\includegraphics[clip=true,width=0.48\textwidth]{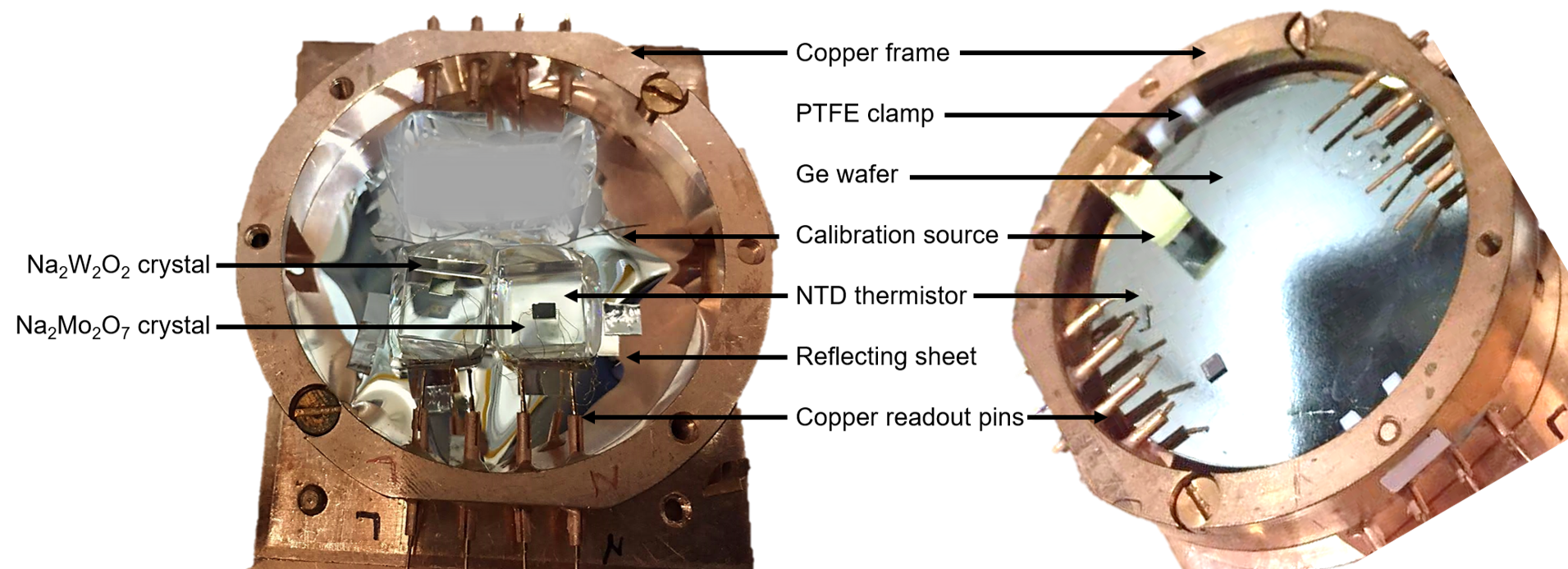}
\hfill
\caption{Picture of the experimental setup, before (left) and after (right) placing the light detector on the Na containing crystals.}
\label{fig:Setup} 
\end{figure}

On the top of the copper frame surrounding the crystals, a second frame is placed, hosting the light detector (LD). It consists of a Ge wafer (50.8 mm diameter, 0.2 mm thick), equipped with an NTD thermistor (size $3\times1\times0.5$ mm$^3$) and held in the copper frame through two PTFE clamps. 
The detector is showed in (Fig.~\ref{fig:Setup}).

Finally, in order to  avoid vibrations reaching the detectors, the setup is  mechanically decoupled from  the cryostat by utilizing a two-stage pendulum system~\cite{Beeman:2013zva} and cooled down to about 10 mK in the CUPID R$\&$D cryostat, a $^3$He/$^4$He dilution refrigerator installed deep underground in Hall C of the Laboratori Nazionali del Gran Sasso, Italy. 

Two sources are included inside the mounting for the detector calibration, a $^{55}$Fe spot facing the LD, and a Thoriated Tungsten wire (1$\%$ Th) lying between the crystals.

The NTD thermistors are biased with a steady current through large (27+27 or 2+2 G$\Omega$) load resistors~\cite{Arnaboldi:2002dzn}.
When the thermistor resistance changes because of an energy release in its absorber, a voltage variation happens across the current-biased thermistor, which is amplified by the front-end electronics located just outside the cryostat~\cite{Arnaboldi:2004jj}.

The signals are then filtered by a 6-pole low pass Bessel-Thomson filter (roll-off rate of 120 dB/decade) with a cut-off frequency of 63 Hz (200 Hz for the LD), and fed into an 18 bit NI-6284 PXI ADC unit. The sampling frequency is 1 kHz for all the channels, while the length of the acquisition window is optimized for each detector according to the different time-development of the signals in each crystal, i.e., Na$_{2}$W$_{2}$O$_{7}$ (1 s), Na$_{2}$Mo$_{2}$O$_{7}$ (500 ms), and LD (500 ms). When a trigger fires on a detector, a waveform is acquired and saved on disk for the off-line analysis. Moreover, when a trigger fires for a Na-containing crystal, the corresponding waveform of the LD is also acquired, regardless of its trigger.
More details about the DAQ system can be found in \cite{DiDomizio:2018ldc}.

\section{Performance and results}
\subsection{Data processing}
Each triggered waveform is analyzed off-line to extract the signal amplitude and other important parameters to quantify the detector performance.
To improve the signal-to-noise ratio, we apply the optimum filter~\cite{Gatti:1986cw}, a software algorithm that suppresses the frequency components of the signal most affected by noise. The construction of the transfer function H($\omega$) of this filter requires the knowledge of the noise power spectra N($\omega$), and the ideal response $S(t)$ of each detector:
\begin{equation}
H(\omega) = k\frac{S^*(\omega)}{N(\omega)}e^{i\omega t_M}
\end{equation}
where $S^*(\omega)$ is the Fourier transform of $S(t)$ and $t_M$ is the time at which the filtered signal is maximum. $S(t)$ is obtained by averaging hundreds of high energy signals, in order to suppress the contributions due to stochastic noise. 
To construct the noise power spectra N($\omega$), we randomly acquire noise windows (i.e. with no pulses therein), and average the square module of their discrete Fourier transform in the frequency domain.

Each pulse is Fourier transformed and filtered using H($\omega$) to obtain a more precise evaluation of its amplitude. The optimum filter allows to extract also some parameters related to the signal time-development, such as the rise time and decay time, defined as the time difference between the 90$\%$ and the 10$\%$ of the rising edge, and the time difference between the 30$\%$ and 90$\%$ of the trailing edge, respectively.

Since the light pulses show a worse signal-to-noise ratio compared to the heat pulses, we estimate their amplitude by accounting for the (known) time-delay between heat and corresponding light signal. More details about this algorithm can be found in~\cite{Piperno:2011fp}. The amplitude extracted by the optimum filter is corrected against temperature drifts by exploiting the physics peak with at the highest energy (the 2.6 MeV line for Na$_2$W$_2$O$_7$ and the 238 keV line for Na$_2$Mo$_2$O$_7$).

Finally, the corrected amplitude is converted into energy. The scintillating crystals are calibrated using the known peaks produced by the thoriated tungsten wire that permanently illuminated the devices. Since the large majority of the energy is released in the form of heat, we expect the light signals to be much smaller (from a few keV to tens of keV depending on the light output of the crystal). Thus, the light detector is calibrated using the 5.9 and 6.4 keV peaks produced by the $^{55}$Fe X-rays source placed on top of the Ge crystal (see Fig.~\ref{fig:Setup} right).

The general features of the detectors are summarized in Tab.~\ref{Table:params_crys}.

\begin{table}[hbtp]
\centering
\begin{tabular}{lcccc}

                        & $A_S$         &$\sigma_{baseline}$  &$\tau_{r}$     &$\tau_{d}$\\
                        &[$\mu$V/MeV]   &[keV RMS]            &[ms]           &[ms]   \\
\hline
\hline
Na$_{2}$Mo$_{2}$O$_{7}$  & 231            &0.2                 &9.1            &30      \\
\hline
Na$_{2}$W$_{2}$O$_{7}$  & 842            &0.1                 &7.5            &22      \\
\hline
Ge-LD       & 1171            &0.3                 &2.2            &7.0      \\
\hline
\hline
\end{tabular}
\caption{Parameters of the cryogenic calorimeters. Voltage response of the thermistor for 1 MeV energy deposit in the crystal (before amplification) $A_S$, energy resolution of the detector baseline after applying the optimum filter ($\sigma_{baseline}$), rise ($\tau_r$) and decay ($\tau_d$) times of the pulses.}
\label{Table:params_crys}
\end{table}

\subsection{Results from Na$_2$W$_2$O$_7$}
\begin{figure}[!h]
\centering
\includegraphics[width=0.45\textwidth]{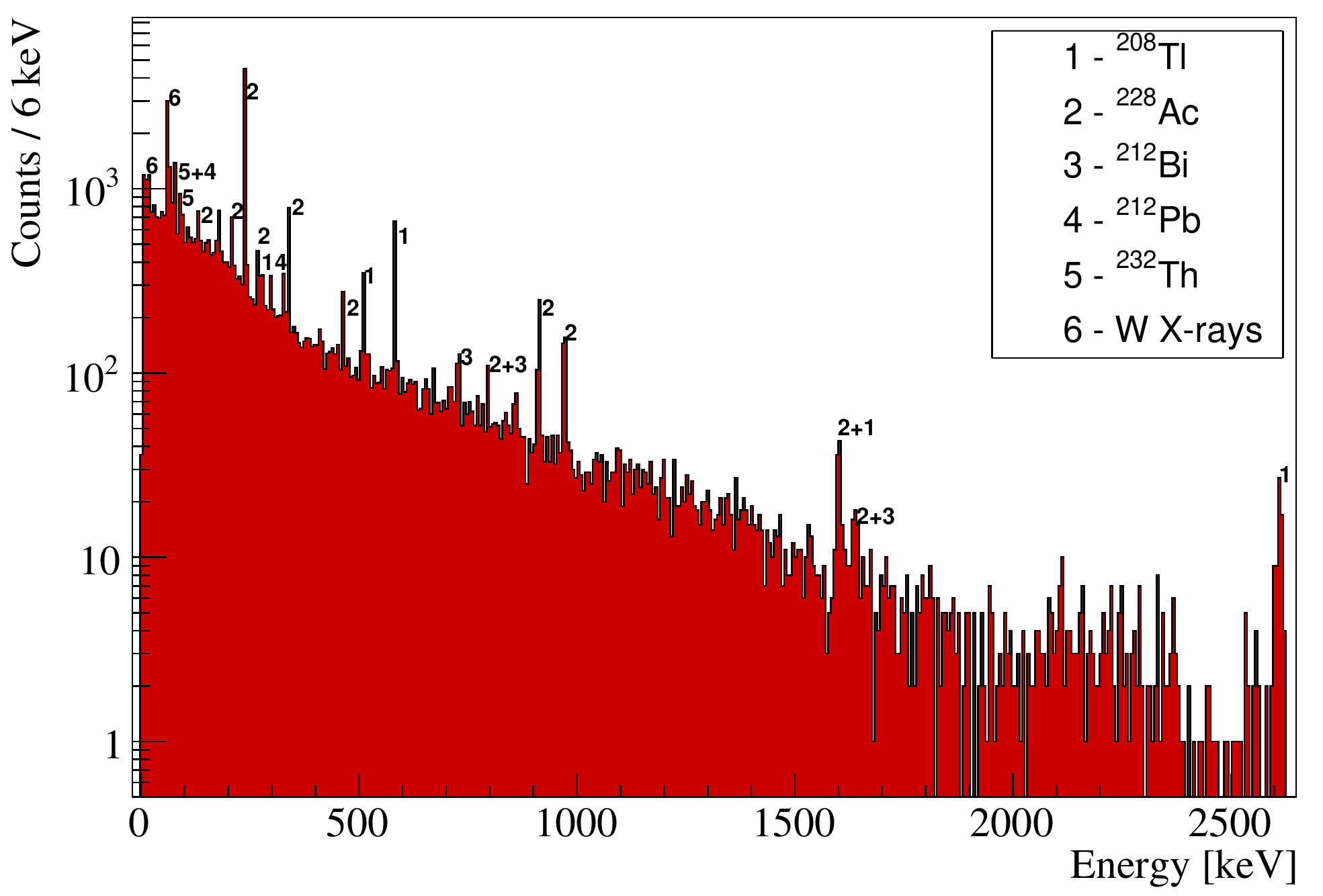}
\caption{Energy spectrum of the Na$_{2}$W$_{2}$O$_{7}$ crystal exposed to the thoriated tungsten wire.}
\label{fig:NaWO_heat} 
\end{figure}

We report in Fig.~\ref{fig:NaWO_heat} the energy spectrum obtained  with the 1~cm$^3$ (5.6.~g)  Na$_{2}$W$_{2}$O$_{7}$ crystal exposed to the thoriated tungsten wire.

\begin{figure}[tb]
\centering
\includegraphics[clip=true,width=0.48\textwidth]{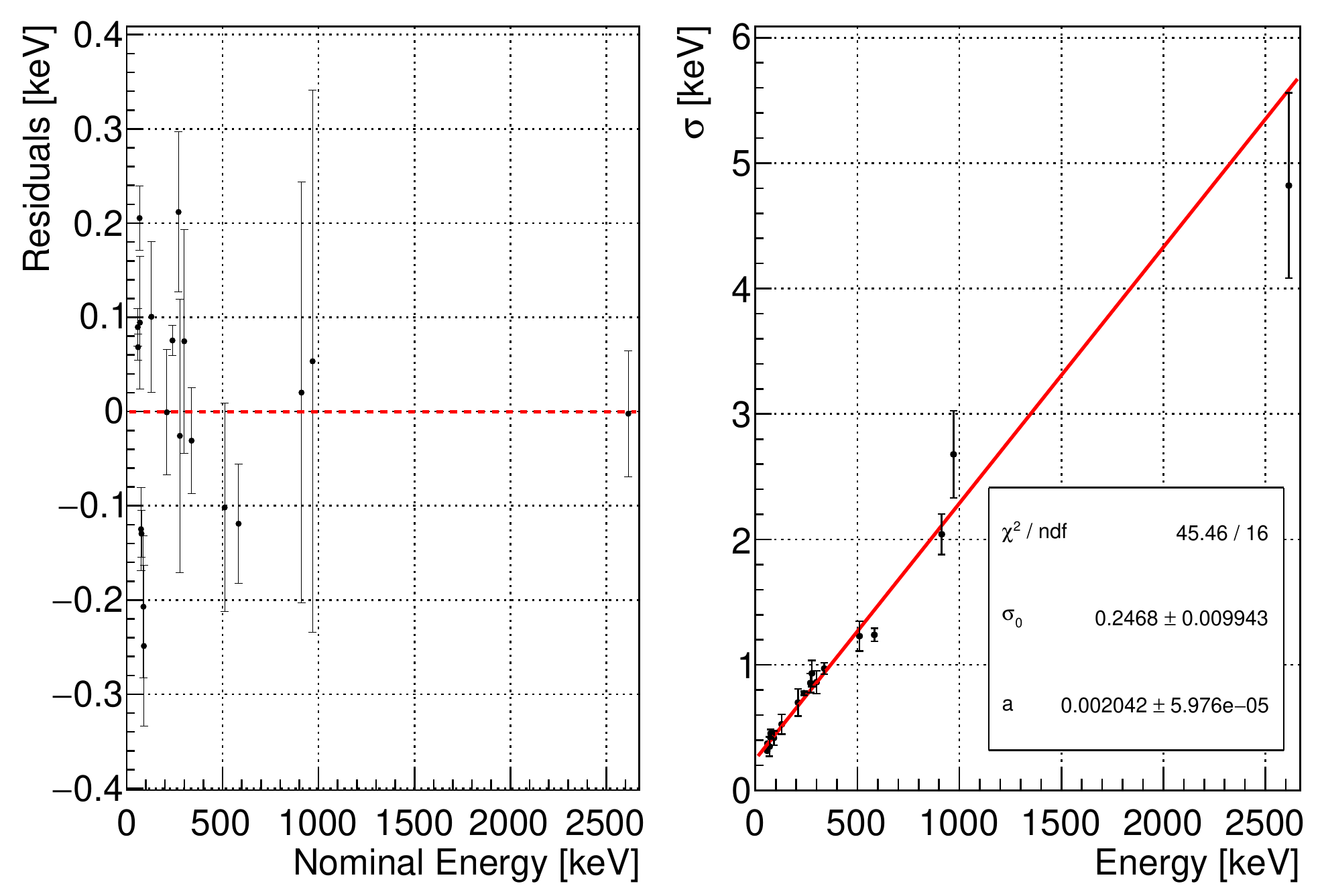}
\hfill
\caption{Left: Residuals of the energy calibration (nominal energy - reconstructed energy) as a function of the energy for Na$_{2}$W$_{2}$O$_{7}$. Right: energy resolution as a function of the energy. We modeled the  dependency on the energy using a linear function $\sigma(E) = \sigma_0 + aE$}
\label{fig:NaWO_calib} 
\end{figure}

The proximity of the $\gamma$-source to the Na$_{2}$W$_{2}$O$_{7}$ detector, as well as the presence of tungsten in the crystal (that increases the probability of photoelectric effect), allow to detect a large number of peaks, from 58 keV (W X-ray escape) up to 2615 keV ($^{208}$Tl). 
We fit each of the 20 detected peaks with a Gaussian function to derive the energy resolution in function of the energy, that increases linearly, as shown in Fig.~\ref{fig:NaWO_calib} (right).

To energy-calibrate this detector, we use a fourth-degree polynomial function  obtaining residuals smaller than 200 eV in the whole considered energy range and better than 100 eV for the vast majority of peaks (Fig.~\ref{fig:NaWO_calib} (left)). 

\begin{figure}[!h]
\includegraphics[width=0.9\linewidth]{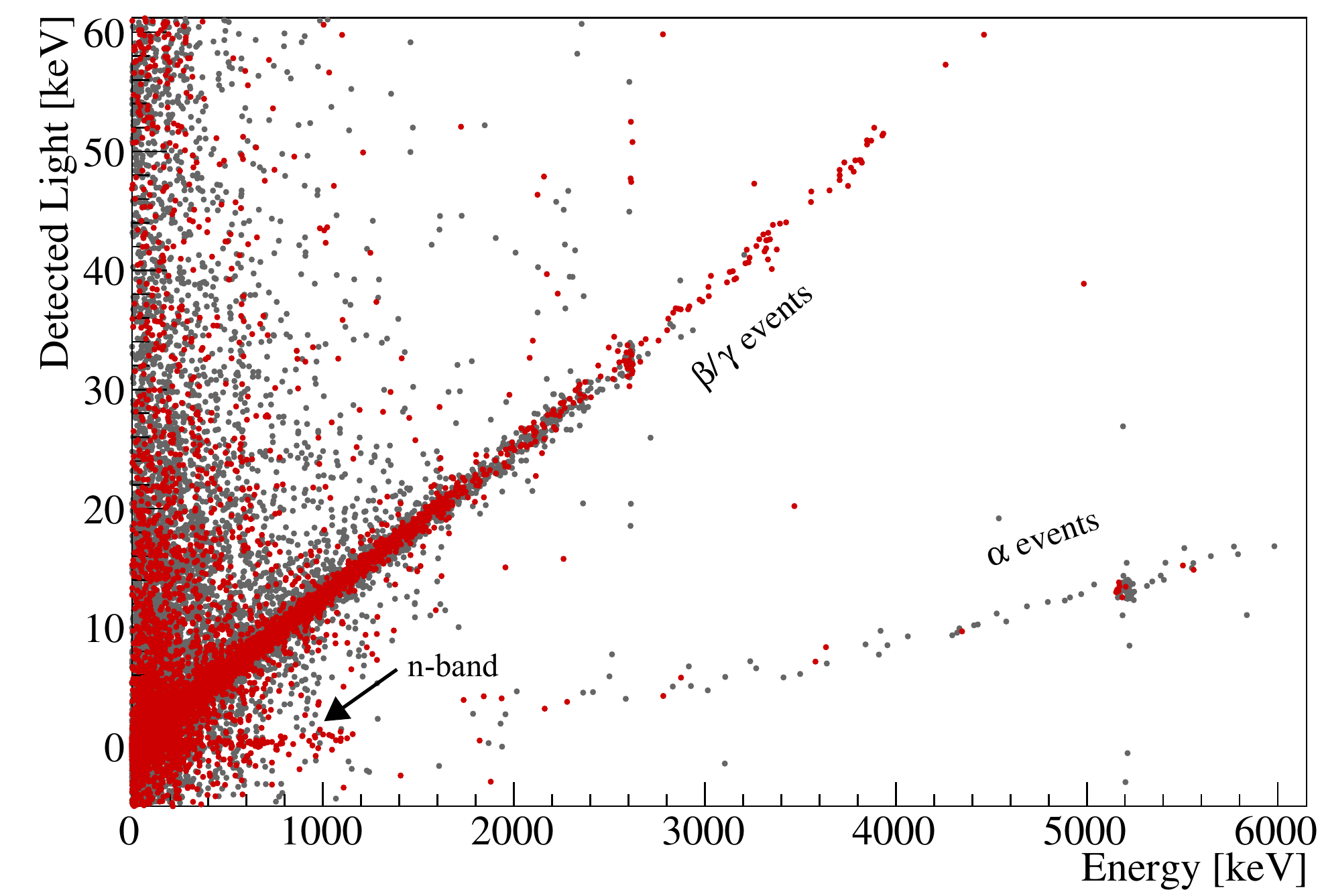}
\caption{Detected light as a function of the heat measured with the Na$_{2}$W$_{2}$O$_{7}$ detector. Gray: run with the thoriated tungsten wire. Red: run with the thoriated tungsten wire and the AmBe neutron source. The nuclear recoils region extends up to $\approx$1 MeV  and is highlighted  by the black arrow.}
\label{fig:NaWO_light_heat} 
\label{fig:all_w2}
\end{figure}

We report in Fig.~\ref{fig:NaWO_light_heat} the detected light as a function of the heat 
signal obtained with the use of a AmBe neutron source placed outside the 
cryostat. We can distinguish three populations, ascribed to $\beta/\gamma$, $\alpha$'s and neutron events.

The light output of $\beta/\gamma$ events increases linearly with the energy and was evaluated with a linear fit up to 5 MeV to include high energy $\gamma$'s emitted during the run with the AmBe source.

The light yield (LY), defined as the measured light for a MeV energy deposit, results  LY$_{\beta/\gamma} = 12.80\pm0.1$ keV/MeV.

For a quick comparison with $\beta/\gamma$ events, we report the LY$_{\alpha}$ measured using the only visible $\alpha$ peak, i.e. the 5304 keV line of $^{210}$Po, LY$_{\alpha}(^{210}$Po) $= 2.61\pm0.1$ keV/MeV. The light Quenching Factor (QF)  of alphas (evaluated at 5.3 MeV) is therefore 
0.204 $\pm$0.008, in good agreement with other crystal 
compounds~\cite{Tretyak:2013xta}.

Finally, we investigated the energy region of nuclear recoils, that appear when the detector is exposed to the neutron source. The energy resolution of the LD prevents an accurate analysis of the light emitted by these interactions, on which however we can put an upper limit of Light$_{nucl.}<$260\,eV.

\subsection{Results from Na$_{2}$Mo$_{2}$O$_{7}$}
\begin{figure}[!h]
\centering
\includegraphics[width=0.48\textwidth]{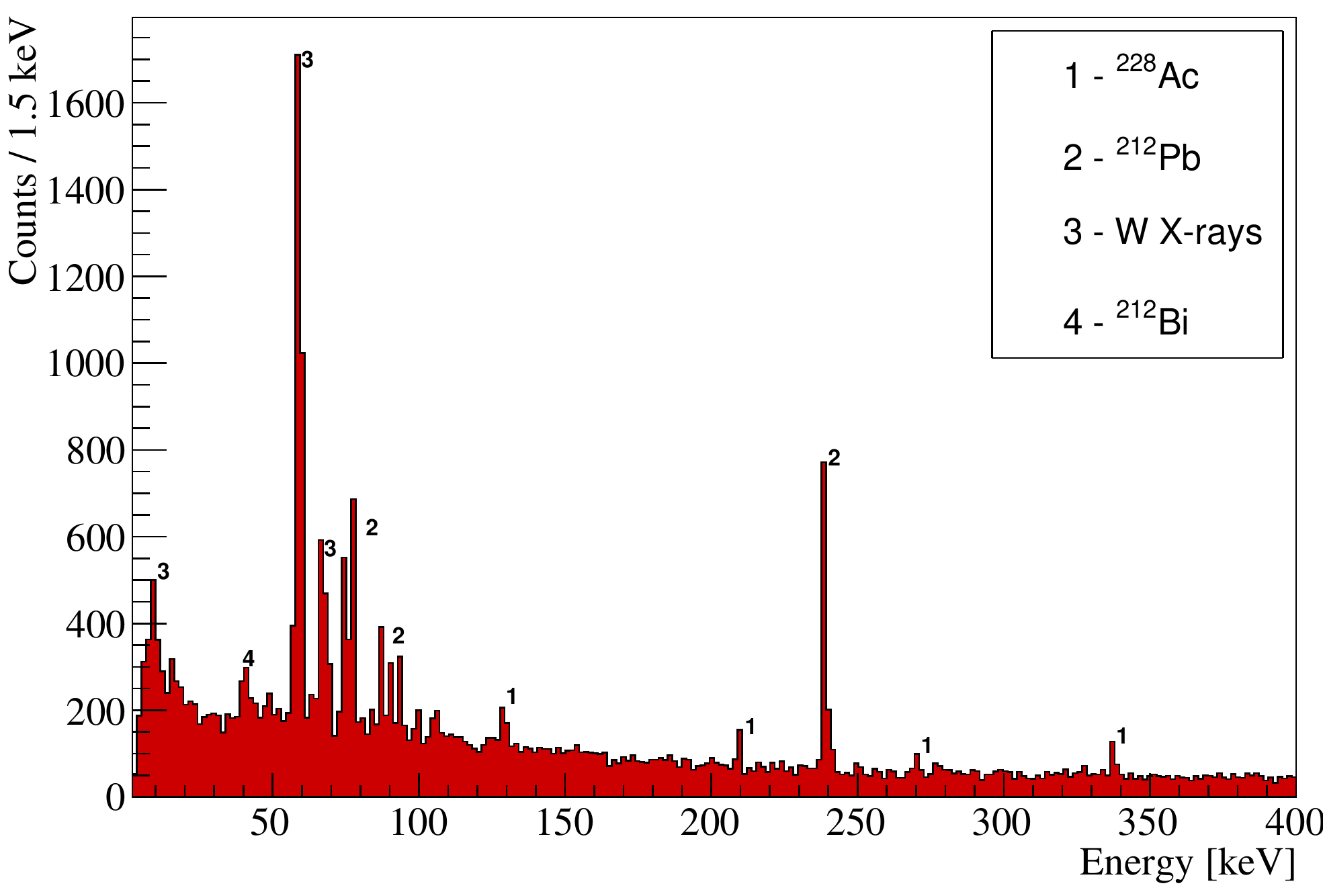}
\caption{Energy spectrum of the Na$_{2}$Mo$_{2}$O$_{7}$ crystal 
exposed to the thoriated tungsten wire.}
\label{fig:NaMoO_heat} 
\end{figure}

Due to the lower density and smaller photoelectric cross-section, the  1 cm$^3$ (3.6.~g) Na$_{2}$Mo$_{2}$O$_{7}$ crystal is less prone to show high energy lines, the largest one being the $\sim$239 keV peak of $^{212}$Pb (Fig.~\ref{fig:NaMoO_heat}).

\begin{figure}[tb]
\centering
\includegraphics[clip=true,width=0.48\textwidth]{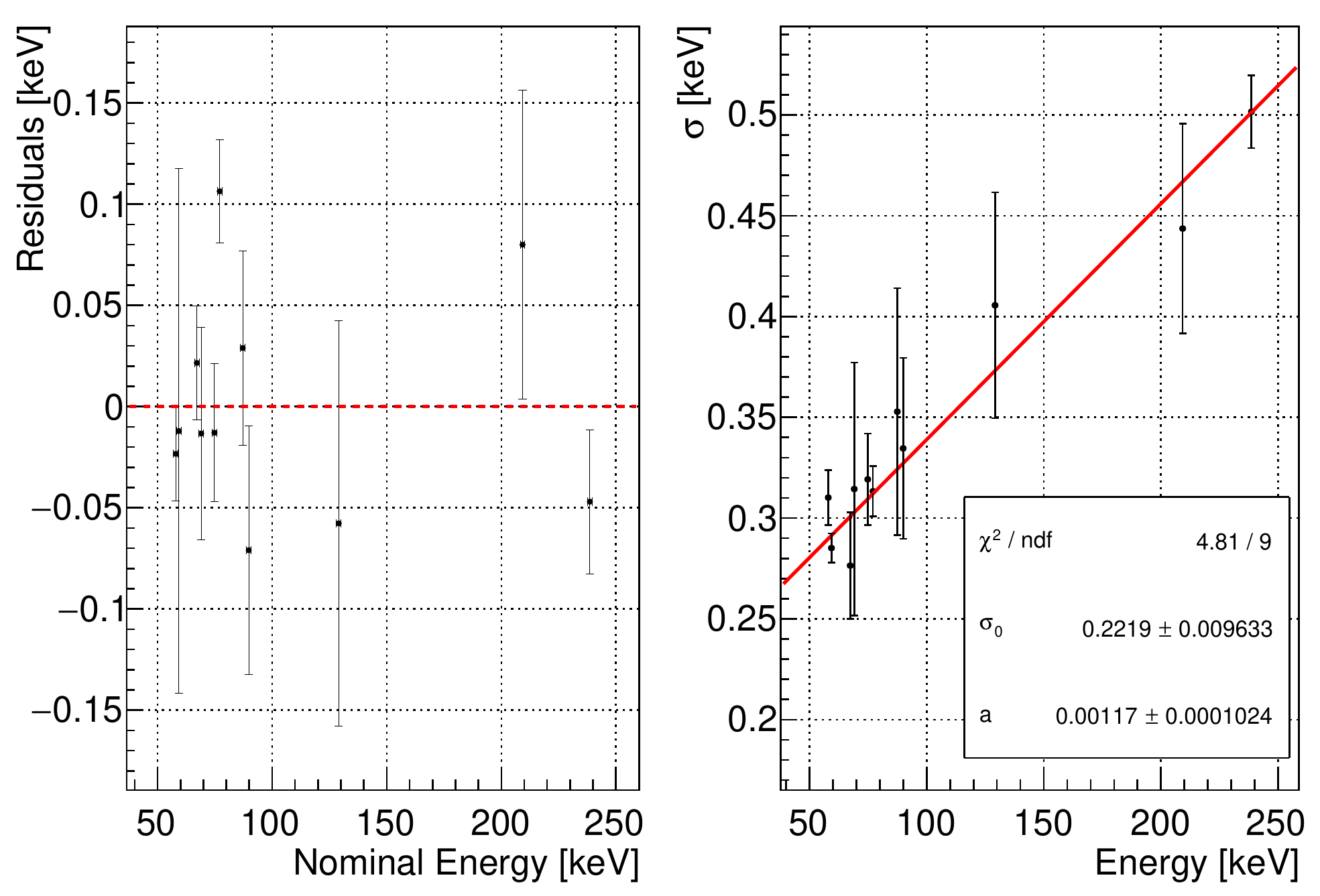}
\hfill
\caption{Left: Residuals of the energy calibration (nominal energy - reconstructed energy) as a function of the energy for Na$_{2}$Mo$_{2}$O$_{7}$. Right: energy resolution as a function of the energy. We modeled the  dependency on the energy using a linear function $\sigma(E) = \sigma_0 + aE$.}
\label{fig:NaMoO_calib} 
\end{figure}

We energy-calibrate the spectrum by applying the same procedure detailed before. The calibration is done with a third degree polynomial  function  and shows residuals smaller than 100 eV in the whole energy range of interest (Fig.~\ref{fig:NaMoO_calib} (left)).

Also in this case, the energy resolution of the device, shown in Fig.~\ref{fig:NaMoO_calib}(right), increases with the energy and can be modelled as $\sigma(E) = \sigma_0 + aE$, with $\sigma_0 = 222\pm 9$ eV and $a = 1.17 \pm 0.10$. 

This detector shows a much smaller LY compared to the one based on  Na$_{2}$W$_{2}$O$_{7}$: LY$_{\beta/\gamma} = 1.61\pm0.01$ keV/MeV and LY$_{\alpha} = 0.25\pm0.02$ keV/MeV.
Due to the poor light yield, it is not possible to disentangle nuclear recoils from the band of $\beta/\gamma$ and thus to estimate their light emission.

We highlight that the presence of $^{100}$Mo, a golden candidate for neutrino-less double beta decay searches, makes this compound very interesting also for a multi-physics experimental approach. Indeed, in the last years the community of double beta decay searches performed an intense R$\&$D activity devoted to the development of crystals containing $^{100}$Mo \cite{Cardani:2013dia,Beeman:2012jd,Armengaud:2017hit,Nagorny:2017wvc,Pattavina:2019eox}.

We extract the energy resolution in the region of interest for this process ($Q_{value} = 3034.40(17)$ keV \cite{Rahaman:2007ng}), obtaining $\sigma$($^{100}$Mo) = 3.77 $\pm$ 0.31 keV. This result is similar to those obtained with other promising compounds, proving the high potential of this crystal  that, unlike Li$_2$MoO$_4$ selected for the CUPID experiment~\cite{Armengaud:2019loe, CUPIDInterestGroup:2019inu}, is not hygroscopic.   

\section{Conclusion and perspectives}
In this paper, we presented the first bolometric measurement of two Na-containing crystals, Na$_2$Mo$_2$O$_7$ and Na$_2$W$_2$O$_7$. 

Despite the low purity of some of the starting materials, the internal contamination measured with ICPMS are extremely low, suggesting a very small segregation factor of impurities in these compounds.

Both the crystals show a high signal response and a baseline energy resolution of the order of $\approx$200 eV  RMS. These preliminary results are very encouraging, as they would guarantee the 1 keV energy threshold required to reach a sensitivity comparable to the one of DAMA/LIBRA.

The energy resolution of these prototypes increases linearly as a function of the energy but remains excellent for both the compounds. This feature is of particular interest for Na$_2$Mo$_2$O$_7$, which could enable the simultaneous search of dark matter interactions and neutrino-less double beta decay of $^{100}$Mo. In the region of interest for this process, indeed, we measured $\sigma=(3.77\pm0.31)$ keV, which is compatible with other Mo-containing bolometers.

We made a preliminary study of the light yield, proving that both the crystals show satisfactory particle identification capabilities at high energies. The high pile-up rate in the light detector (due to the presence of the -very intense- thoriated tungsten wire) prevented a more detailed study of the light yield of nuclear recoils, on which we could just set upper limits. 

This pilot experiment proves that both the compounds should be further studied for applications in rare events searches.

\section*{Acknowledgements}
We thank the CUPID-0 and the CUORE collaborations for the overall support and for sharing their DAQ and software. We  express our gratitude to LNGS for the generous hospitality  and, in particular, we want to thank the LNGS mechanical workshop team E. Tatananni, A. Rotilio, A. Corsi, and B. Romualdi for continuous and constructive help and M. Guetti for his cryogenic expertise and his constant support.
\bibliography{main}
 \bibliographystyle{ieeetr}
 
 \end{document}